\documentstyle[aps]{revtex}

\input boxedeps
\HideDisplacementBoxes
\ForceWidth{7.8cm}
\ForceOn

\SetRokickiEPSFSpecial          

\begin{document}
\draft
\twocolumn[\hsize\textwidth\columnwidth\hsize\csname @twocolumnfalse\endcsname

\title{Einstein--Yang--Mills Black Hole Interiors: \\
       Serious Problems But Simple Solution}
\author{Mikhail Zotov}
\address{Nuclear Physics Institute, Moscow State University,
         Moscow 119899, Russia}  
\date{\today}
\maketitle

\begin{abstract}
 Recently E.~E.~Donets, D.~V.~Gal'tsov, and the author reported
 the results of numerical and analytical investigation
 of the  SU(2) Einstein--Yang--Mills (EYM) black hole interior
 solutions  (gr-qc/9612067). It was shown that a generic interior
 solution develops a new type of an infinitely oscillating behavior
 with exponentially growing amplitude. Numerical data  for three
 sequential oscillations were presented. The numerical integration
 technique was not discussed. Later P.~Breitenlohner, G.~Lavrelashvili,
 and D.~Maison confirmed our main results (gr-qc/9703047). But they
 have made some misleading statements. In particular, they claimed,
 discussing the oscillations, that ``as one performs the numerical
 integration one quickly runs into serious problems...'' so that
 ``it is practically impossible to follow more than one or two of
 them numerically'' because ``the numerical integration procedure
 breaks down'' (pp.~3, 12). It is shown here that trivial logarithmic
 substitutions and integration along the integral curve solve
 these ``serious problems'' easily.
\end{abstract}
\pacs{04.20.Jb, 97.60.Lf, 11.15.Kc, 02.60.Jh}
]
\narrowtext

\vspace*{-1mm}

\begin{flushright}
 {\it ``Well'', said Owl, ``the customary \\ procedure
       in such cases is as follows.'' \\
       A. A. Milne ;)}
\end{flushright}

\vspace{7mm}


 Recall the main notation from \cite {we}. We
 assume the static spherically symmetric magnetic ansatz for the
 YM potential
\[
             A = \left(W(r)-1\right) \left(T_{\varphi}d\theta-
                 T_{\theta}\sin\theta d\varphi\right) \, ,
\]
 ($T_{\varphi, \theta}$ are
 spherical projections of the $\rm SU(2)$ generators),
 and the following parametrization of the metric:
\[
     ds^2 = (\Delta/r^2) \sigma^2dt^2-(r^2/\Delta) dr^2-r^2d\Omega^2 \, ,
\]
 where $d\Omega^2 = d\theta^2 + \sin^2\theta d\varphi^2$, and
 $\Delta$, $\sigma$ depend on $r$.
 The field equations include a coupled system for $W$, $\Delta$:
\begin{eqnarray}
                 && \Delta U' + F W' = W V/r \, ,  \label{eqW}  \\
                 && (\Delta/r)' +  2 \Delta U^2 =F \, , \label{eqDelta}
\end{eqnarray}
 where $U = W'/r$,
 $V=W^2-1$, $F=1 - V^2/r^2$, and a decoupled equation for $\sigma$:
\[
                       (\ln\sigma)' = 2rU^2 \, .
\]
 These equations admit BH solutions in the domain
 $r\ge r_h$ for any  radius $r_h$ of the event horizon.

 The system (\ref{eqW}, \ref{eqDelta}) was integrated numerically
 in the region $0<r<r_h$ using an
 adaptive step size Runge--Kutta method
 starting at the left vicinity of the event horizon $r_h$ with one
 free parameter $W_h = W(r_h)$ satisfying inequalities  $|W_h|<1$ and
 $1-W_h^2<r_h$, which are the necessary conditions for asymptotic flatness.

 One of the main results of our investigation
 is that  a generic EYM BH interior solution
 develops a new type of an infinitely oscillating behavior
 with exponentially growing amplitude, during  which the `falls'
 of $\Delta$  ($\Delta < 0$) may be approximated as
\begin{equation}
     \Delta(r) = \frac{\Delta(r_k)}{r_k}\; r \;
                 \exp\left[U(r_k)^2 (r^2_k-r^2)\right], \quad r_k \ge r \, ,
\label{Dexp}
\end{equation}
 where $r_k$ is the ``starting point'' of the $k$-th oscillation,
 $(U(r_k) r_k)^2 \gg 1$, and all the regime can be described by a
 two-dimensional dynamical system (see \cite{we} for the details).


 Our investigation of the EYM BH interior solutions
 \cite{we,MSc,Odessa,ITEP,Novgorod}
 reveals, that  one meets only two problems during
 numerical integration of (\ref{eqW}, \ref{eqDelta}).

 {\it Problem 1.\/}
 Very small values of $r$ as $r \rightarrow 0$ and $|\Delta|$ in the
 neighborhoods of $\Delta$'s local maxima, and very large values of $|U|$
 as $r \rightarrow 0$,  and $|\Delta|$ in the neighborhoods of 
 $\Delta$'s local minima.

 {\it Problem 2.\/}
 Intervals of very slow variation of $r$ but extremely fast
 variation of $\Delta$ and maybe $U$.

 Obviously, the first problem can hardly be called a problem
 because it may be solved by trivial substitutions like
 $\ln(1/r)$ (or just $\ln r$), $\ln \ln (1/r)$ , etc., for $r$;
 $\ln (-\Delta)$, $\ln |\ln (-\Delta)|$, etc., for $\Delta$ an  so on.

 The second one is not much more difficult. Let us show how it was
 solved for obtaining numerical data for three sequential oscillations,
 presented in \cite{we} (we shall also use figures 6, 7 from \cite{we}).

\begin{figure}[t]
 \begin{center}
   {\BoxedEPSF{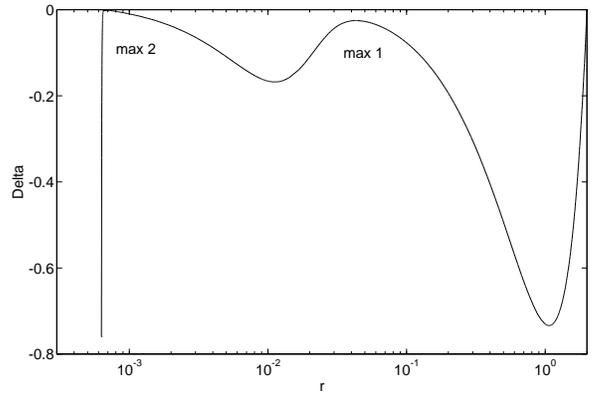}}
 \end{center}
 \caption{The beginning of $\Delta$-oscillations  for 
          $r_h=2$, $W(r_h)=-0.342072$ ($n=1$ BH solution).}
 \label{fall}
\end{figure}
\begin{figure}[htbp]
 \begin{center}
   {\BoxedEPSF{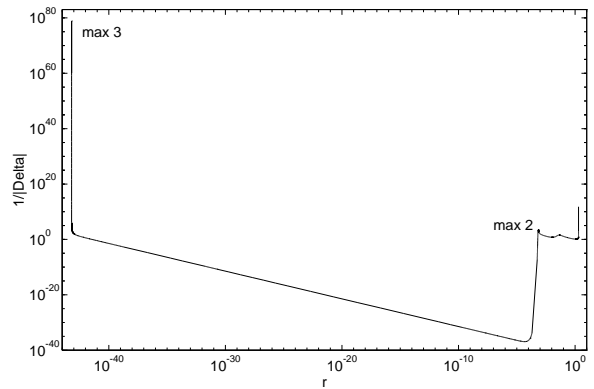}}
 \end{center}
 \caption{$|\Delta|^{-1}$ for the second oscillation.}
 \label{osc2}
\end{figure}
\begin{figure}[htbp]
 \begin{center}
   {\BoxedEPSF{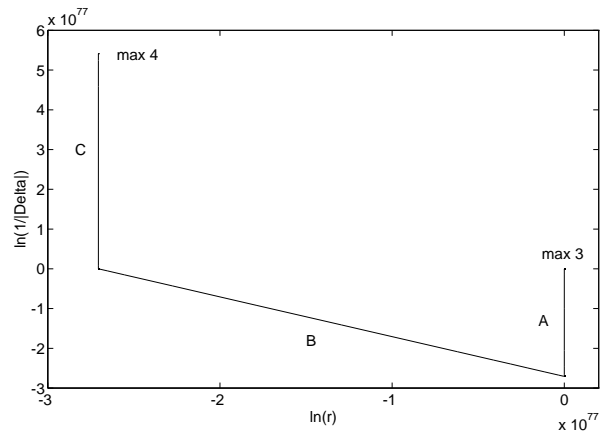}}
 \end{center}
 \caption{$\ln(|\Delta|^{-1})$ for the third oscillation.}
 \label{osc3}
\end{figure}

 Let us rewrite (\ref{eqW}, \ref{eqDelta}) in a more ``comfortable''
 way, suggested in \cite{VG89}:
\begin{eqnarray}
         && DW'' + FW' = \frac{1}{x}VW \, ,  \label{eqWx} \\
         && D' + \left(\alpha W'^2-\frac{1}{2x}\right) D = F \, , \label{eqD}
\end{eqnarray}
 where $x = (r/r_h)^2$, $D = 2\Delta/r_h^2$, $\alpha=4/r_h^2$,
 $V = (W^2 - 1)/2$, $F = 1 - \alpha V^2/x$,
 and $' \equiv d/dx$. Here $W'$ plays the same role as $U$ in (\ref{eqW},
 \ref{eqDelta}).
 In a normal form, needed for the Runge--Kutta method, this system reads as
\begin{eqnarray}
 &&  W' = P \, , \nonumber \\
 &&  P' = \frac{1}{xD} V (W + \alpha V P) - \frac {P}{D}\,, \label{normal}\\
 &&  D' = F - \left(\alpha P^2-\frac{1}{2x}\right)D\,. \nonumber
\end{eqnarray}
 It is clear from (\ref{normal}) that small (absolute) value of
 denominator $xD$, which takes place in the metric function local
 maximum neighborhood,
 may cause numerical problems. It may be unessential for some
 first maxima (`{\sf max~1}' and `{\sf max~2}' in Fig.~\ref{fall}),
 but becomes considerable as oscillations progress. Really,
 integrating system (\ref{normal}) for $r_h = 2$, $W_h = -0.342072$,
 one finds out that $x$ practically stops in the third local maximum
 neighborhood (`{\sf max~3}' in Figs.~\ref{osc2}, \ref{osc3}) with
 $D \approx -10^{-16}$, $W' \approx 10^{18}$ (I mean a PC with 15 digits
 after the decimal point). This does not allow to pass the third maximum
 correctly.
%
%
 One of the possible solutions is  to desingularize (\ref{normal})
 by introducing a parameter $t$ as
\begin{equation}
                  dx = xD dt.
\label{xD}
\end{equation}
 But this method is
 in some sense local: it allows to pass the local maximum,
 but does not allow to reach the following minimum of $D$.
 More general technique is based on integration  along 
 the integral curve.    


 Recall the basic formula for the curve length (see, e.g., \cite{mathan}
 or any other textbook on integral calculus).
 Let $\Gamma$ be a smooth curve, defined by
\[
 x = \phi(t), \quad y = \psi(t), \quad z = \chi(t), \quad t \in [a, b],
\]
 where $\phi$, $\psi$, $\chi \in C^1[a, b]$, and
 $\phi'^2 + \psi'^2 +\chi'^2 > 0$ at $[a, b]$.
 Then the length $L$ of $\Gamma$ is expressed as
\[
            L = \int_a^b \sqrt{\phi'^2 + \psi'^2 +\chi'^2} dt.
\]
 This simple formula allows to solve our numerical problems
 by introducing a parameter, representing the length of the
 integral curve, and to perform numerical integration 
 along the integral curve instead of the radial coordinate $x$ (or $r$).
 This allows to pass the intervals of slowly varying $x$ easily.


 Thus, one may introduce a parameter, say $l$, such that
 $dl = (W'^2 + P'^2 + D'^2)^{1/2} dx$ (or, for example,
 $dl = (1 + P'^2)^{1/2} dx$  to integrate along $P$, etc.)
 and to rewrite system (\ref{normal}) in a relevant way.
 But we will introduce $t = \ln x$ first in order to have the solution
 components of comparable values, and $G = \ln(-D)$ in order to pass not only
 the third local maximum, but also the third local minimum (see (\ref{Dexp})
 and Fig.~\ref{osc3}).
 Thus, the new system may be written as
\begin{equation}
 \left[
  \begin{array}{c}
                  \dot t \\
                  \dot W \\
                  \dot P \\
                  \dot G
  \end{array}
 \right] = \frac{dt}{dl}
 \left[
  \begin{array}{l}
                  1 \\
                  {\rm e}^t P \\
                  NP - QVW \\
                  \frac{1}{2} - \alpha {\rm e}^t P^2 - N
  \end{array} \right],
 \label{f}
\end{equation}

\vspace{3ex}
\vspace{3ex}
\begin{figure}[t]
 \begin{center}
   {\BoxedEPSF{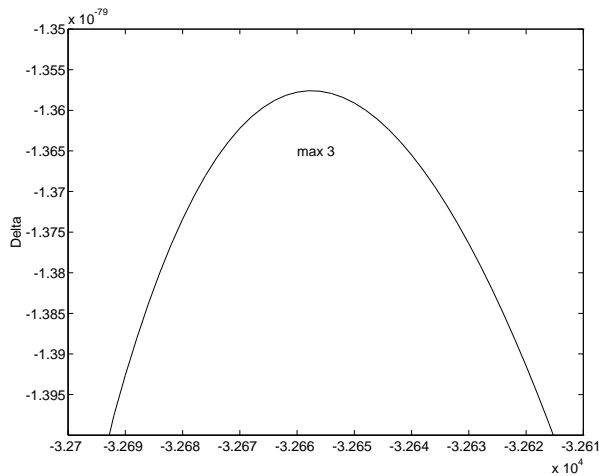}}
 \end{center}
 \caption{$\Delta$ versus $l$: the third local maximum.}
 \label{max3}
\end{figure}

\noindent
 where $Q = \exp(-G)$, $H = {\rm e}^t - \alpha V^2$,
 $N = HQ$, $\dot { } \equiv d/dl$,  and, for example,
 $dl = (1 + {\dot G}^2)^{1/2} dt$ for integration along $G$.
 This system allows to pass the third local maximum  and parts
 {\sf A} and {\sf B} of the third oscillation (Figs.~\ref{osc3}, 
 \ref{max3}).

\noindent
 Note that part  {\sf B} in fact does not need to be passed along the
 integral curve. It is sufficient just to use $t = \ln x$ (or, e.g.,  
 $t = \ln(1/x)$) as an independent variable.

 At last, one needs to pass part {\sf C} (Fig.~\ref{osc3}) in order
 to reach (and pass) the fourth local maximum of $\Delta$.
 $\Delta$ vanishes fast at this interval; this forces $W'$
 to increase strongly
 ($|U \Delta| \approx |U(r_3^{max})|$ at this interval \cite{Novgorod}).
 Therefore  one needs to introduce $Z = \ln W'$. System
 (\ref{f}), rewritten in the corresponding way, solves this problem.
 (Evidently, this can be done before passing `{\sf max~3}', but
 (\ref{f}) works a bit better along part {\sf A}.)


 It is also convenient to improve a system, prepared for integration
 along part {\sf C} in a way similar to (\ref{xD}).
 Let us introduce another parameter, say $p$,
 such that $dl = -D dp$. Then one obtains
\[
 \left[
  \begin{array}{c}
                  \dot p \\
                  \dot t \\
                  \dot W \\
                  \dot Z \\
                  \dot G
  \end{array}
 \right] = \frac{dp}{dl}
 \left[
  \begin{array}{l}
                  1 \\
                  E \\
                  E \exp (t+Z) \\
                  H-VW \; \exp(-Z) \\
                  \frac{1}{2} E - H - \alpha E \; \exp(t+2Z)
  \end{array} \right],
\]
 where $E = \exp(G)$.
 This system works fine both along part {\sf C} and while passing
 local maxima (both `{\sf max~3}' and `{\sf max~4}').

 The next oscillation needs the next ``order'' of logarithmic
 substitutions ($\ln \ln(1/x)$, etc.), but the numerical integration
 technique remains the same.

\begin{figure}[t]
 \begin{center}
   {\BoxedEPSF{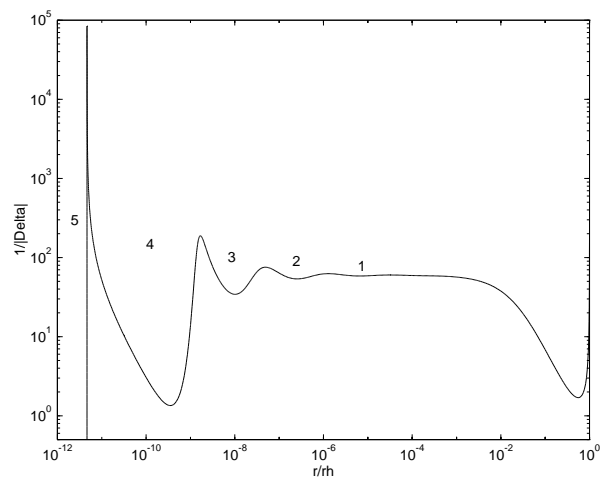}}
 \end{center}
 \caption{This curve represents a solution, singular in the exterior
          region. One can see 4 first oscillations of $\Delta$ and the
          beginning of the 5th one, which can be easily 
          integrated using the technique, described above.}
 \label{5osc}
\end{figure}

 Recall, at last, that the investigation, presented in \cite{we},
 was restricted to the class of asymptotically flat solutions.
 If one also studies solutions, which are singular in the exterior region,
 then a situation, shown in Fig.~\ref{5osc} is rather typical. This
 makes the claims cited in the abstract even more surprising.

 Thus, if we do not discuss restrictions, which are inherent 
 for every numerical method,
 there are hardly any ``serious problems'', which
 can ``break down'' the numerical integration of the SU(2) EYM equations
 inside the event 
 horizon. All the difficulties, connected with the numerical
 investigation, are typical and can be solved by  integration along
 the integral curve.

 I would like to thank  Rostislav Zhukov for his little 286 PC.
 The research was supported in part by the RFBR grant 96-02-18899.


\end{document}